# Evaluation of the Design Metric to Reduce the Number of Defects in Software Development

M. Rizwan Jameel Qureshi
*Faculty of Computing & Information Technology of King Abdul Aziz University, Jeddah, Saudi Arabia*
Email: anriz@hotmail.com

Waseem A. Qureshi
*Senior Executive Manager, Corporate Operations Division, Banque Saudi Fransi, Head Office, Riyadh 11554, Saudi Arabia*
Email: waseem1e@gmail.com

*Abstract*—Software design is one of the most important and key activities in the system development life cycle (SDLC) phase that ensures the quality of software. Different key areas of design are very vital to be taken into consideration while designing software. Software design describes how the software system is decomposed and managed in smaller components. Object-oriented (OO) paradigm has facilitated software industry with more reliable and manageable software and its design. The quality of the software design can be measured through different metrics such as Chidamber and Kemerer (CK) design metrics, Mood Metrics & Lorenz and Kidd metrics. CK metrics is one of the oldest and most reliable metrics among all metrics available to software industry to evaluate OO design. This paper presents an evaluation of CK metrics to propose an improved CK design metrics values to reduce the defects during software design phase in software. This paper will also describe that whether a significant effect of any CK design metrics exists on total number of defects per module or not. This is achieved by conducting survey in two software development companies.

## 1. Introduction

The demand of quality software is increasing day-by-day due to social dependency of the clients on the software. For instance architecture, interface and integration etc are the main software design defects. Any problem in software can cause financial loss and time delays. Today's software must assure consistent and error free execution whenever it is used. Software design has an important role in the quality of the software. Poor design will result in greater rework and higher cost [1]. Design defects need to be identified in early stages of system development life cycle (SDLC). There are significant research studies showing that defect seeding at the design phase are visible in the maintenance phase [2]. Different technologies provide different facility to improve the quality of the design. A lot of research has been done on different metrics to assure the quality of software design [1]. So it is very important to have a good software design to reduce the maintenance time and overall cost of project.

Different Metrics are available to evaluate software design quality. The data, about Chidamber and Kemerer (CK) Metrics and total number of defects, is collected from two software companies/houses to conclude the results. The main objective of this paper is to propose a framework to quantitatively evaluate software design and observe its effects on total number of defects.

The remainder of this paper is organized as follows: Section 2 illustrates the related work. Section 3 discusses the CK metrics. Section 4

presents motivation for hypothesis. Section 5 describes the research question and hypothesis. Section 6 presents the research setting. Section 7 provides the findings of experiments and analysis. Conclusion is given in the final section.

## 2. Related Work

The object oriented (OO) approach to software development assures better management of software complexity and a likely improvement in project outcomes in terms of quality and timelines. There has been a lot of research on metrics for OO software development in recent years, which shows that OO methodology and project outcomes have some relationship [3]. In the OO environment, certain integral design concepts such as inheritance, coupling, and cohesion have been argued to significantly affect complexity [4].

The concepts of software metrics are well established, and many metrics relating to product quality have been developed and used. To evaluate a metric's usefulness as a quantitative measure of software quality, it must be based on the measurement of a software quality attribute. Software metrics plays an important role to improve requirement engineering, design quality, code quality, reducing overall defects in the SDLC phases and product readiness to ship/deploy. For example, one study recently showed that prediction models using design metrics had an error rate of only 9% when estimating the proportion of classes with post-release defects for a commercial Java application [5]. This is encouraging because such estimates can be used to allocate maintenance resources and for obtaining assurances about software quality. Another study estimated corrective maintenance cost savings of 42% by using OO metrics [6]. Here, classes containing defects were predicted early in the project and were targeted for inspection.

Chidamber and Kemerer proposed first suite of OO design measures that is called as CK Metrics [4]. The authors of this suite of metrics claim that these measures can aid users in understanding design complexity, in detecting design flaws and in predicting certain project outcomes and external software qualities such as software defects, testing, and maintenance effort. CK Metrics helps to analysis complexity, understandability / usability, reusability / application specific and testability-/maintainability. Thus it is important to have a metrics program in all the phases of SDLC to observe the quality of the input, process and output.

## 3. CK Metric

The complete details of the CK metrics [7] along with the names, common names and definition are given in the following Table 1. The information in the Table 2 is gathered through an Internet survey [8]. Table 2 shows that these values are provided by the different software developed by different vendors for "Metrics calculation domain". It is also interesting to know that these values as mentioned in Table 2 differ from each other, proving that there is no unanimous cut point threshold value for these metrics. These vendors however, have not provided any basis of these proposed values.

## 4. Motivation for Research Question and Hypothesis

Software design plays an important role in the development of software. Software design describes how the system is decomposed and organized into components. Metric is a mean to quantitatively evaluate quality. CK, Mood and Lorenz & Kidd metrics are discussed in the literature [9]. CK metrics is one of the most popular OO design metrics and hence there is no need to compare this metrics with others. Someone can construct this CK metrics manually, but there are tools available to do this job. Automated tools and process has significant edge over the manual process in terms of time, efficiency and accuracy. The authors surveyed and found Together-Soft, SD-Metrics and Objecteering tools which are contributing significantly in the industry to measure the design quality. These tools are used by the industrial giants like Sun Microsystems, Microsoft etc. The literature survey shows:
1. That CK metrics being the most used is the most trusted and popular of all the metrics.
2. That there is no, agreed upon, cut point threshold value of CK metrics that industry uses.
3. That it depends upon the historical data of the organization as to what values suit for the organization [10].

Table 1: CK Metrics

|   | Metric Name | Definition |
|---|---|---|
| 1 | Weighted Methods per class (WMC) | This measure is an aggregate count of the number of methods in each class. This count also includes Constructors and Destructors of the class. |
| 2 | Depth Of Inheritance Tree (DIT) | This count is the maximum length / depth from the node to the root of the tree. |
| 3 | Number of immediate subclasses (NOC) | Number of children / subclasses subordinated to a class in the class hierarchy. |
| 4 | Coupling between Objects Classes (CBO) | It is a count of the number of other classes to which it [a class] is coupled. |
| 5 | Response for a class (RFC) | It is a count of the set of methods that can potentially be executed in response to a message received by an object of that class. |
| 6 | Lack of cohesion in Method (LCOM) | It is a count of the number of method paired whose similarity is 0 minus the count of method paired whose similarity is not 0. |

Table 2: Threshold Values For CK Metrics By Different Vendors/Researchers

| Sr. # | CK Metrics | Rosenber, NASA | SD-Metrics | Together Soft | Objecteering Enterprise Edition | Cantata++ |
|---|---|---|---|---|---|---|
| 1 | WMC | 40 | - | 100 | 3-7 | - |
| 2 | DIT | 6 | 0-3 | 4 | 0-4 | - |
| 3 | NOC | - | - | - | 1-4 | - |
| 4 | CBO | 5 | 0-31 | 30 | 1-4 | - |
| 5 | RFC | 100 | 3-365 | - | - | - |
| 6 | LCOM | - | - | - | - | - |

According to Caper Jones [2], "Defect seeding at the design phase is visible in the maintenance phase". CK metrics being a means of reducing defects in design phase and hence in maintenance phase, it is therefore important to find out an improved/unanimous version of CK-metrics' values which is being done in this research work.

The research question can now be set forth in the next section on the basis of literature review.

## 5. Research Question and Hypotheses

How to evaluate CK design metrics to reduce the number of defects in software development? Following hypotheses are used in this research.

**H₀ (Null Hypothesis)** There is no relationship between CK Metrics [WMC, DIT, NOC, CBO, RFC, and LCOM] and the total number of defects found per module of software system.
H0: $\mu1 \neq \mu2 \neq \mu3 \neq \mu4 \neq \mu5 \neq \mu6$.

**H₁ (Alternate Hypothesis)** There is a relationship between CK metrics [WMC, DIT, NOC, CBO, RFC, and LCOM] and the total number of defects found per module of the software system.
H1: $\mu1 = \mu2 = \mu3 = \mu4 = \mu5 = \mu6$.

## 6. Research Setting

The research site for data collection is two leading software development companies as given in the following Table III. The core competencies of first software company include all areas of the Internet technologies, client/server applications, object-oriented technologies, groupware automation and large

Evaluation of the Design Metric to Reduce the Number of Defects in Software Development

Table 3: Organization Details

| Organization Details | |
|---|---|
| Organization Size | 1st Software Company: 1000+ employees<br><br>2nd Software Company: 500+ Employees |
| Organization's Maturity level | 1st Software Company: CMM Level 5, ISO 9001<br><br>2nd Software Company: None |
| Project Details | |
| Projects under Experiment | 1st Software Company Project: P1<br><br>2nd Software Company Projects: P2, P3, P4<br>Three projects with 12 modules in total. |
| Domain of the Projects Under Study | 1st Software Company Project/P1: Leasing<br>2nd Software Company Project/P2: Web Portal including financial packages |
| Duration of the Projects | P1: 12 Months<br>P2: 18 Months<br>P3: 12 Months<br>P4: 06 Months |
| Team Size | 1st Software Company<br>1. Project Manager = 1<br>2. Architect = 1<br>3. Analyst = 2<br>4. Developers = 5<br>*[Analyst and also involved in development]*<br>5. QA persons = 3<br><br>2nd Software Company<br>1. Project Manager = 1<br>2. Architect = 1<br>3. Analyst = 1<br>4. Developers = 6<br>*[Analyst and also involved in development]*<br>5. QA persons = 2 |
| Technology Used | Java/ J2EE/SQL Server |
| SDLC Followed | Tailored Waterfall methodology |
| Average Experience of Team | Medium |

scale system integration. It's a CMM Level 5 company. Total strength of the company is more than 1000 employees in total. It is the first Pakistani software development company who achieved CMM level 5. This is the main reason for selection of this company. The 2nd software company is also a leading provider of real-time financial portal software. The company is based at Chicago. The company has been building real-time financial portal technology dating back to 1998, which leverages the Internet for the aggregation of real-time data, news and applications. They have developed financial portals utilizing data from Reuters, MarketWatch, Barcharts, Money.net, Edgar-Online, S&P, Zacks, Hyperfeed, Morningstar, Briefing.com and many others. Rest of the details is provided in Table III. It may be mentioned that both of the software companies are using tailored Waterfall methodology as mentioned in the Table 3. By changing the development methodology, the results of the CK design metrics may also influence with the results of that methodology.

Following points also need attention of the readers to know little bit more for the software companies.

1. Due to length issue of research paper, the authors are not attaching questionnaire used for the survey regarding the research presented in this paper. They do have the questionnaire for the reference and verifications of those whoever is needed.
2. The authors can not disclose the names of the IT companies those have been

surveyed for this paper. This is because the companies have participated in the survey subject to the condition that their names will not be disclosed.

Different types of research methodologies exist, in today's research world, depending upon the nature of research problem. As far as the research methodology of this paper is concerned, survey is used to collect the data.

A team of two-liaison persons from local 1st software company & 2nd software company were dedicated to assist in data collection and verification. The team includes one person from software quality assurance (SQA) department and the other from development department.

Automated tools and processes have significant edge over the manual processes in terms of time and efficiency. Due to this reason, a survey is made using Internet to find the available automated tools to measure design quality. Borland Together Edition for JBuilder Version 6.1 is used to calculate metrics from the code. It is used by the industrial giants like Sun Microsystems and Microsoft Corporation [11].

Regression analysis will be used to test the hypotheses. The purpose of regression analysis is to develop a predictive model that could predict the number of defects for a module in a similar environment [discussed in the later part of this paper]. The model will estimate the number of defects regardless of their nature, based on the historical data available, using multiple regression analysis. In this case:

| | | |
|---|---|---|
| **Dependent Variable** | = | Total number of defects per module |
| **Independent Variables** | = | WMC, DIT, NOC, CBO, RFC, LCOM |

**Null Hypothesis**

$H_0$: None of the independent variables has a significant effect on the dependent variable.

$H_0$: $\beta j = 0$ *(Where j= 1,2,3,4,5,6)*

**Alternate Hypothesis:**
$H_1$: At least one of the independent variables has a significant effect on the dependent variable.

$H_1$: $\beta j \neq 0$ *At least for one value of j (where j= 1,2,3,4,5,6)*

## 7. Experiment and Analysis

Seven days on the average, 5 to 6 hours, have been spent to collect data, its verification and validation in each company. For data verification, code is inspected manually to make sure different metrics have the correct data. Then randomly some classes are selected to validate and verify the data gathered by Together-Soft.

This study is concerned with the number of total defects only. Unfortunately the defects segregated by their origin could not be found for example requirements, design, and coding. Due to tight deadlines, companies could not invest time in further categorizing defects with respect to their origin and severity level. This study is also not focusing on the severity levels of the defects. The selected projects are from the same implementation domain [J2EE] and having at least 2 to 2.5 years of experience of each member in the project. So any one can fairly assume that there exist some design problems in the total number of defects and not all the defects are of low severity. The authors have further verified this by manually going through the bug report and found that some of the defects were of high severity and were tracing back to the design.

Table 4 shows data that is analyzed to calculate CK metrics. Graphs are plotted with modules on x-axis and CK metrics [CBO, DIT, LCOM, NOC, RFC and WMC] on y-axis separately as given in the following figures from Figure 1 to Figure 6 respectively. The average and threshold values are plotted on the graphs. By plotting these lines, one can clearly see three regions in the graphs.

1. Below the lower plotted straight line
2. Between two straight lines
3. Above the upper straight line.

Calculations are made for three regions by using the following formulation.

Total Number of defects from module below lower limit = X
Total Number of independent variable from module below lower limit = Y
Defects per independent variable = X/Y     (1)

The straight lines are plotted by using minimum, average and maximum values if no threshold values exist for any independent variable. A comparison can be made that in which region minimum numbers of defects are occurring.

On the basis of this analysis, significant region can be identified and accepted as the best among three regions with less number of defects/ independent variable. A summery will be presented in at the end to summarize the finding of this phase.

Linear regression is applied on the data using SPSS tool to generate the results.

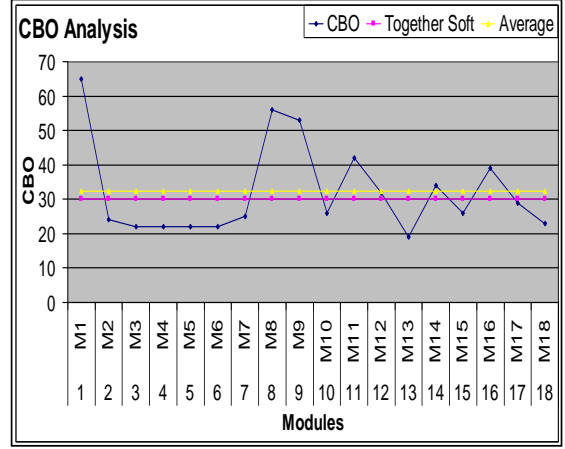

Fig. 1: CBO Analysis

Table 4: Data Collected to Calculate CK Metrics

| Sr. # | Modules | CBO | DIT | LCOM | NOC | RFC | WMC | Total Number of defects | Bug Fixing Time Man hours |
|---|---|---|---|---|---|---|---|---|---|
| 1 | M1 | 65 | 3 | 11223 | 142 | 149 | 577 | 307 | 184 |
| 2 | M2 | 24 | 3 | 12132 | 122 | 289 | 1647 | 111 | 48 |
| 3 | M3 | 22 | 3 | 1276 | 112 | 109 | 564 | 75 | 48 |
| 4 | M4 | 22 | 3 | 11669 | 238 | 287 | 1560 | 186 | 48 |
| 5 | M5 | 22 | 3 | 5048 | 20 | 185 | 1051 | 35 | 160 |
| 6 | M6 | 22 | 3 | 9051 | 37 | 145 | 998 | 66 | 48 |
| 7 | M7 | 25 | 7 | 261 | 21 | 376 | 165 | 35 | 192 |
| 8 | M8 | 56 | 3 | 6832 | 101 | 312 | 651 | 30 | 128 |
| 9 | M9 | 53 | 3 | 1459 | 101 | 196 | 245 | 37 | 96 |
| 10 | M10 | 26 | 5 | 758 | 12 | 148 | 332 | 45 | 192 |
| 11 | M11 | 42 | 5 | 1562 | 90 | 195 | 375 | 23 | 120 |
| 12 | M12 | 32 | 7 | 367 | 18 | 386 | 89 | 35 | 192 |
| 13 | M13 | 19 | 5 | 419 | 10 | 119 | 196 | 47 | 192 |
| 14 | M14 | 34 | 6 | 5470 | 81 | 322 | 594 | 8 | 32 |
| 15 | M15 | 26 | 5 | 758 | 12 | 148 | 332 | 45 | 192 |
| 16 | M16 | 39 | 7 | 2821 | 58 | 425 | 560 | 37 | 96 |
| 17 | M17 | 29 | 6 | 2821 | 47 | 392 | 644 | 81 | 288 |
| 18 | M18 | 23 | 6 | 228 | 8 | 392 | 52 | 16 | 96 |

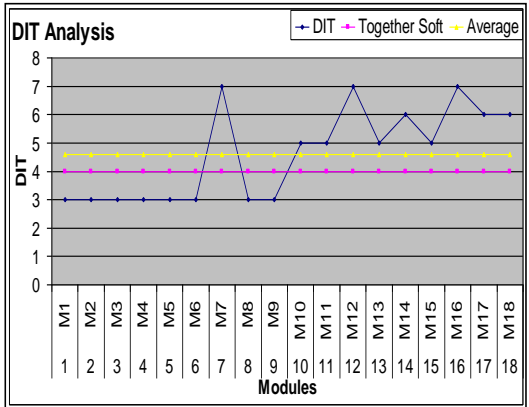
Fig. 2: DIT Analysis

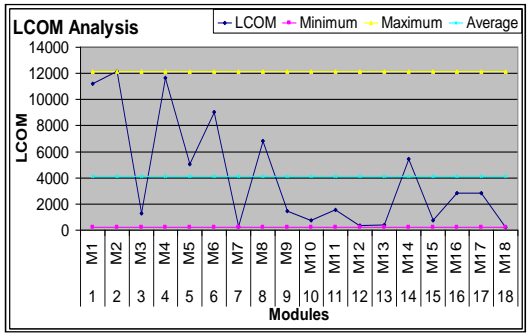
Fig. 3: LCOM Analysis

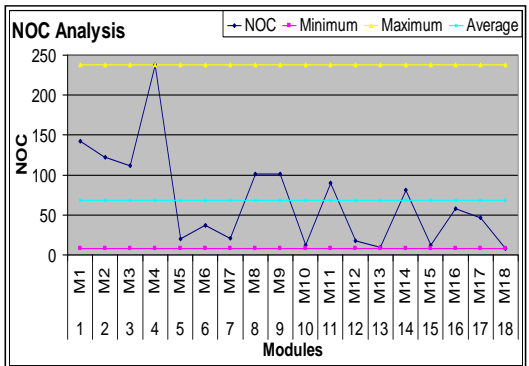
Fig. 4: NOC Analysis

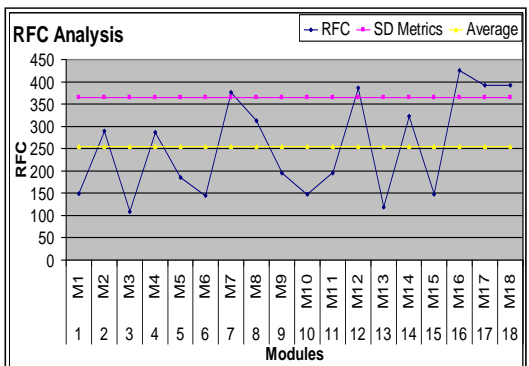
Fig. 5: RFC Analysis

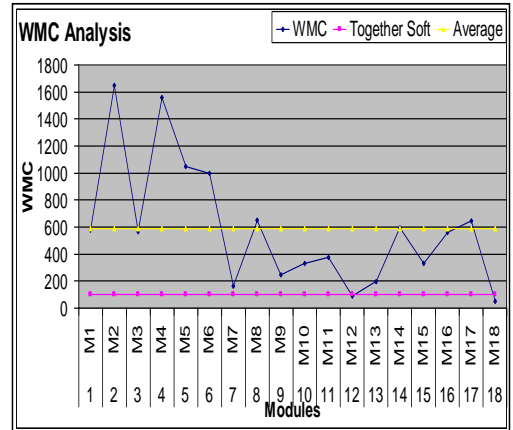
Fig. 6: WMC Analysis

Table 5 shows the analysis of the data that is collected from both companies where survey is conducted. Row 1 of Table V presents the CBO values from the data collected. The average of CBO [which is 32.3] is compared by the average suggested by Together Soft, which is 30, which are almost equal.

Table 5: Analysis of data

|  | Min | Max | Median | Average | Std. Dev |
|---|---|---|---|---|---|
| CBO | 19.0 | 65.0 | 26.0 | 32.3 | 13.5 |
| DIT | 3.0 | 7.0 | 5.0 | 4.6 | 1.6 |
| LCOM | 228.0 | 12132.0 | 2191.5 | 4119.7 | 4280.6 |
| NOC | 8.0 | 238.0 | 52.5 | 68.3 | 61.3 |
| RFC | 109.0 | 425.0 | 241.5 | 254.2 | 110.1 |
| WMC | 52.0 | 1647.0 | 562.0 | 590.7 | 461.5 |
| LOC | 807.0 | 86922.0 | 7577.0 | 16637.7 | 22547.4 |
| Defects | 8.0 | 307.0 | 41.0 | 67.7 | 72.6 |

MODULES HAVE VALUES $0 < CBO < 30$
Total number of defects from M2, M3, M4, M5, M6, M7, M10, M13, M15, M17, M18 = 260

Total number of CBO from M2, M3, M4, M5, M6, M7, M10, M13, M15, M17, M18 = 742

Defects per CBO = 260/742 = 0.35

MODULES HAVE VALUES $30 < CBO < 32.3$
There lie no values in between 30 and 32.3

MODULES HAVE VALUES $32.3 < CBO < 65$
Total number of defects from M1, M8, M9, M11, M12, M14, M16 = 477

Total number of CBO from M1, M8, M9, M11, M12, M14, M16 = 321

Evaluation of the Design Metric to Reduce the Number of Defects in Software Development

Defects per CBO = 477/321= 1.49

The results in Table 5 show that the value of CBO should be 0 < CBO < 30 in order to minimize the number of defects / CBO. Row 2 of Table V presents the DIT values. The Averages of DIT from the data collected [which is 4.6] is compared by the average suggested by Together Soft, which is 4.0.

MODULES HAVE VALUES 0 < DIT < 4
Total number of defects from M1, M2, M3, M4, M5, M6, M8, M9 = 847

Total number of DIT from M1, M2, M3, M4, M5, M6, M8, M9 = 24

Defects per DIT = 847/24 = 35.29

MODULES HAVE VALUES 4 < DIT < 4.61
There are no values in between 4 and 4.61

MODULES HAVE VALUES 4.61 < DIT < 7.0
Total number of defects from M7, M10, M11, M12, M13, M14, M15, M16, M17, M18=372

Total number of DIT from M7, M10, M11, M12, M13, M14, M15, M16, M17, M18=59

Defects per DIT = 372/59 = 6.31

The analysis of data shows that the acceptable value of DIT should be greater than 4.61 and less than 7, in order to minimize the number of defects/DIT. Row 3 of Table V presents the LCOM values from the data collected. The Average of the LCOM is 4119.72, the minimum LCOM is 228.00, and maximum LCOM is 12132.00.

MODULES HAVE VALUES
228.00 < LCOM < 4119.72
Total number of defects from M3, M7, M9, M10, M11, M12, M13, M15, M16, M17, M18=751

Total number of LCOM from M3, M7, M9, M10, M11, M12, M13, M15, M16, M17, M18=12730

Defects per LCOM = 751/12730 = 0.06

MODULES HAVE VALUES
4119.72 < LCOM < 12132.00
Total Number of defects from M1, M2, M4, M5, M6, M8, M14 = 468

Total Number of LCOM from M1, M2, M4, M5, M6, M8, M14 = 61425

Defects per LCOM = 468/61425 = 0.01

The results show that the value of LCOM should be 4119.72 < LCOM < 12132.00 to minimize the number of defects / LCOM. Row 4 of Table V presents the NOC values from the data collected. The Averages of NOC from data collected [which is 68.8] is compared by the average suggested by Objecteering Enterprise Edition, which is 1 for minimum and 4 for maximum [10]. The minimum values from the data collected are '8' with is double than what Objecteering Enterprise Edition suggests.

MODULES HAVE VALUES
8 < NOC < 68.33
Total number of defects from M5, M6, M7, M10, M12, M13, M15, M16, M17, M18 = 617

Total number of NOC from M5, M6, M7, M10, M12, M13, M15, M16, M17, M18= 243

Defects per NOC = 617/243= 2.54

MODULES HAVE VALUES
68.33 < NOC < 238
Total number of defects from M1, M2, M3, M4, M8, M9, M11, M14 = 602

Total number of NOC from M1, M2, M3, M4, M8, M9, M11, M14 = 987

Defects per NOC = 602/987 = 0.61

Table 5 describes that the value of NOC should be 68.33 < NOC < 238.00 to minimize the number of defects/NOC. The graph of Fig.5 presents the RFC values from the data collected. The Averages of RFC [which is 254.17] is compared by the maximum suggested by SD Metrics, which is 365 [10].

MODULES HAVE VALUES
0 < RFC < 254.17
Total number of defects from M1, M3, M5, M6, M9, M10, M11, M13, M15 = 680

Total number of RFC from M1, M3, M5, M6, M9, M10, M11, M13, M15 = 1394

Defects per RFC = 680/1394 = 0.48

MODULES HAVE VALUES
254.7 < RFC < 365

Total number of defects from M2, M4, M8, M14=335

Total number of RFC from M2, M4, M8, M14 = 1210

Defects per RFC = 335/1210 = 0.28

MODULES HAVE VALUES
365 < RFC < 425
Total number of defects from M7, M12, M16, M17, M18 = 204

Total number of RFC from M7, M12, M16, M17, M18 = 1971

Defects per RFC = 204/1971 = 0.1

The results in Table 5 suggest that the value of RFC should be 365 < RFC < 425 to minimums the number of defects/RFC. The graph of Fig.6 presents the WMC values. The average of WMC [which is 590.67] is compared by the maximum suggested by Together Soft, which is 100. It is also interesting to know that the total number of defects from M1 to M6 is greater than the total number of defects from M7 to M18 which is in between the compared values.

MODULES HAVE VALUES
0 < WMC <100
Total number of defects from M12, M18 = 51

Total number of WMC in M12, M18 = 141

Defects per WMC = 51/141 = 0.36

MODULES HAVE VALUES
100 < WMC < 590.67
Total number of defects from M1, M3, M7, M9, M10, M11, M13, M15, M16 = 651

Total number of WMC from M1, M3, M7, M9, M10, M11, M13, M15, M16 = 3346

Defects per WMC = 651/3346= 0.05

MODULES HAVE VALUES
590.7 < WMC < 1647
Total number of defects from M2, M4, M5, M6, M8, M14, M17 = 517

Total number of WMC from M2, M4, M5, M6, M8, M14, M17 = 7145

Defects per WMC = 517/7145 = 0.07

The results in Table 5 advises that the value of WMC should be $100 < WMC < 590.67$ for low defects/WMC. Table 6 shows the summary of the results.

Table 6: Summary of the Results

| Sr. # | CK Metrics | Findings | Defect |
|---|---|---|---|
| 1 | CBO | 0<CBO<30 | 0.35/CBO |
| 2 | DIT | 4.61<DIT< 7 | 6.31/DIT |
| 3 | LCOM | 4119.72 < LCOM < 12132.00 | 0.01/LCOM |
| 4 | NOC | 68.33<NOC<238.00 | 0.61/NOC |
| 5 | RFC | 365<RFC<425 | 0.1/RFC |
| 6 | WMC | 100<WMC<590.67 | 0.05/WMC |

Table 7: Summary of the Model

| Model | R | R Square | Adjusted R Square | Std. Error of the Estimate |
|---|---|---|---|---|
| 1 | .830(a) | .688 | .519 | 50.36631 |

Table 8: Results of ANOVA Test

| ANOVA (b) | | | | | | |
|---|---|---|---|---|---|---|
| Model | | Sum of Squares | df | Mean Square | F | Sig. |
| 1 | Regression | 61671.189 | 6 | 10278.532 | 4.052 | .022(a) |
| | Residual | 27904.422 | 11 | 2536.766 | | |
| | Total | 89575.611 | 17 | | | |
| a. Predictors: (Constant), WMC, RFC, CBO, NOC, DIT, LCOM | | | | | | |
| b. Dependent Variable: Defects | | | | | | |

Table 7 shows the value of $R^2$ that is 0.688 indicating that 68.8% of the variation in dependent variable is explained by the independent variables. The value of '$R^2$' is 0.688 indicates that 68.8% of the variation in dependent variable is explained by the independent variables in the linear regression.

The 'F' value in Table 8 shows variance of data indicating the significance of the derived model. The authors find the LCOM (form the t-values in Table 9 for the individual regression coefficients) is the only metrics that has a significant effect on the total number of defects. The remaining factor contributes insignificantly.

Table 9: Individual Regression Coefficients

| Model | | Unstandardized Coefficients | | Standardized Coefficients | t | Sig. |
|---|---|---|---|---|---|---|
| | | B | Std. Error | Beta | | |
| 1 | (Constant) | 32.803 | 103.195 | | .318 | .757 |
| | CBO | -.652 | 1.545 | -.121 | -.422 | .681 |
| | DIT | 13.956 | 17.528 | .310 | .796 | .443 |
| | LCOM | .020 | .008 | 1.155 | 2.588 | .025 |
| | NOC | .556 | .316 | .469 | 1.760 | .106 |
| | RFC | -.239 | .191 | -.362 | -1.246 | .239 |
| | WMC | -.113 | .084 | -.715 | -1.343 | .206 |
| A Dependent Variable: Defects | | | | | | |

a) Predictors: (Constant), WMC, RFC, CBO, NOC, DIT, LCOM

**Regression Equation**

$$Y = a + \beta_1 X_1 + \beta_2 X_2 + \beta_3 X_3 + \cdots + \beta_n X_n \quad (2)$$

$$Y = 32.803 - (0.121*CBO) + (0.31*DIT) + (1.155*LCOM) + (0.469*NOC) - (0.362*RFC) - (0.715*WMC)$$

## 8. Conclusion

It is important to evaluate quality while designing software. CK metrics helps to evaluate design quality. The regression analysis shows that all the independent variables [CBO, DIT, NOC, WMC, RFC] have an insignificant effect on the total number of defects except LCOM. LCOM is the only attribute that has a significant effect on the total number of defect. Rest of the independent variables bears a significant effect on the total number of defects, hence the null hypothesis is rejected and the alternative hypothesis is accepted. Software development companies should concentrate on LCOM to control the design defects. Time for bug fixing is also collected. Once we have predicted the total number of defects, we can easily calculate the time required for bug fixing.

**M. Rizwan Jameel Qureshi:** Assistant Professor of Faculty of Computing & Information Technology in King Abdul Aziz University, Saudi Arabia interested in software engineering and database systems.

**Waseem Qureshi:** Senior Executive Manager, Corporate Operations Division, Banque Saudi Fransi, Head Office, Riyadh 11554, Saudi Arabia.